\documentclass[12pt,times]{article}
\title{PPINN: Parareal Physics-Informed Neural Network for time-dependent PDEs}

\author{Xuhui Meng$^{1*}$,
        Zhen Li$^{2}$\footnote{The first two authors contributed equally to this work.}~,
        Dongkun Zhang$^1$
        and George Em Karniadakis$^1$\footnote{Corresponding Email: \href{mailto:george_karniadakis@brown.edu}{george\_karniadakis@brown.edu}} \\
\small{$^1$~Division of Applied Mathematics, Brown University, Providence, RI 02912, USA}\\
\small{$^2$~Department of Mechanical Engineering, Clemson University, Clemson, SC 29634, USA}}

\date{\today}

\RequirePackage{lineno}
\usepackage[colorlinks=true]{hyperref}
\usepackage{graphicx}
\usepackage{multirow,array,color,mathrsfs}
\usepackage{subfigure}
\usepackage{epstopdf}
\usepackage{caption,cite,float}
\usepackage{algpseudocode}
\usepackage[ruled,vlined]{algorithm2e}

\DeclareGraphicsExtensions{.eps,.mps,.pdf,.jpg,.png}
\graphicspath{{figures/}{../figures/}}

\usepackage{fancyhdr}
\usepackage{amsmath}

\usepackage{enumitem}

\usepackage{pxfonts}

\usepackage[top=1.5cm,bottom=2.5cm,left=1.5cm,right=1.5cm]{geometry}

\usepackage[figuresright]{rotating}
\usepackage{extarrows}

\usepackage{fixltx2e,xspace}

\usepackage{amsmath,amssymb}
\begin{document}

\maketitle
\vspace{-20pt}
\begin{abstract}
\noindent
Physics-informed neural networks (PINNs) encode physical conservation laws and prior physical knowledge into the neural networks, ensuring the correct physics is represented accurately while alleviating the need for supervised learning to a great degree~\cite{2019Raissi}. While effective for relatively short-term time integration, when long time integration of the time-dependent PDEs is sought, the time-space domain may become arbitrarily large and hence training of the neural network may become prohibitively expensive. 
To this end, we develop a {\em parareal} physics-informed neural network (PPINN), hence decomposing a long-time problem into many independent short-time problems supervised by an inexpensive/fast coarse-grained (CG) solver. 
In particular, the serial CG solver is designed to provide approximate predictions of the solution at discrete times, while initiate many fine PINNs simultaneously to correct the solution iteratively.
There is a two-fold benefit from training PINNs with small-data sets rather than working on a large-data set directly, i.e., training of individual PINNs with small-data is much faster, while training the fine PINNs can be readily parallelized. Consequently, compared to the original PINN approach, the proposed PPINN approach may achieve a significant speedup for long-time integration of PDEs, assuming that the CG solver is fast and can provide reasonable predictions of the solution, hence aiding the PPINN solution to converge in just a few iterations. 
To investigate the PPINN performance on solving time-dependent PDEs, we first apply the PPINN to solve the Burgers equation, and subsequently we apply the PPINN to solve a two-dimensional nonlinear diffusion-reaction equation.
Our results demonstrate that PPINNs converge in a couple of iterations with significant speed-ups proportional to the number of time-subdomains employed. 
\end{abstract}

\section{Introduction}\label{sec:int}
At a cost of a relatively expensive computation in the training process, deep neural networks (DNNs) provide a powerful approach to explore hidden correlations in massive data, which in many cases are physically not possible with human manual review~\cite{2018Hodas}.
In the past decade, the large computational cost for training DNN has been mitigated by a number of advances, including high-performance computers~\cite{2018Kurth}, graphics processing units (GPUs)~\cite{2019Lew}, tensor processing units (TPUs)~\cite{2019You}, and fast large-scale optimization schemes~\cite{2018Bottou}, i.e., adaptive moment estimation (Adam)~\cite{2014Kingma} and adaptive gradient algorithm (AdaGrad)~\cite{2011Duchi}. In many instances for modeling physical systems, physical invariants, e.g., momentum and energy conservation laws, can be built into the learning algorithms in the context of DNN and their variants~\cite{2019Michoski,2019Doan,2019Mattheakis}. This leads to a physics-informed neural network (PINN), where physical conservation laws and prior physical knowledge are encoded into the neural networks~\cite{2019Raissi,1994Dissanayake}. Consequently, the PINN model relies partially on the data and partially on the physics described by partial differential equations (PDEs).

\begin{figure}[ht!]
  \centering
  \includegraphics[width=0.6\textwidth]{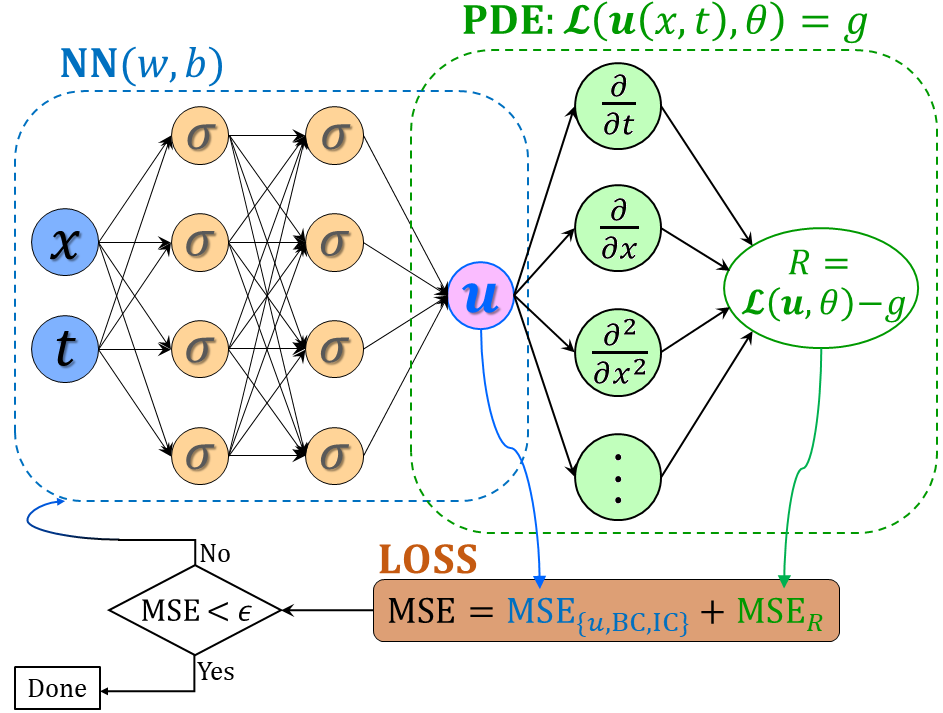}
  \caption{Schematic of a physics-informed neural network (PINN), where the loss function of PINN contains a mismatch in the given data on the state variables or boundary and initial conditions, i.e., ${\rm MSE}_{\{u, {\rm BC, IC}\}} = N_u^{-1}\sum||\mathbf{u}(x,t)-\mathbf{u}_\star||$, and the residual for the PDE on a set of random points in the time-space domain, i.e., ${\rm MSE}_R = N_R^{-1}\sum||R(x,t)||$. The hyperparameters of PINN can be learned by minimizing the total loss ${\rm MSE}={\rm MSE}_{\{u, {\rm BC, IC}\}}+ {\rm MSE}_R$.
  }
  \label{fig:PINN}
\end{figure}

Different from traditional PDE solvers, although a PDE is encoded in the neural network, the PINN does not need to discretize the PDE or employ complicated numerical algorithms to solve the equations. Instead, PINNs take advantage of the {\em automatic differentiation} employed in the backward propagation to represent all differential operators in a PDE, and of the training of a neural network to perform a nonlinear mapping from input variables to output quantities by minimizing a loss function. In this respect, the PINN model is a grid-free approach as no mesh is needed for solving the equations. All the complexities of solving a physical problem are transferred into the optimization/training stage of the neural network. Consequently, a PINN is able to unify the formulation and generalize the procedure of solving physical problems governed by PDEs regardless of the structure of the equations. Figure~\ref{fig:PINN} graphically describes the structure of the PINN approach~\cite{2019Raissi}, where the loss function of PINN contains a mismatch in the given data on the state variables or boundary condition (BC) and initial condition (IC), i.e., ${\rm MSE}_{\{u,{\rm BC, IC}\}} = N_u^{-1}\sum||\mathbf{u}(x,t)-\mathbf{u}_\star||$ with $\mathbf{u}_\star$ being the given data, combined with the residual of the PDE computed on a set of random points in the time-space domain, i.e., ${\rm MSE}_R = N_R^{-1}\sum||{R}(x,t)||$. Then, the PINN can be trained by minimizing the total loss ${\rm MSE}={\rm MSE}_{\{u, {\rm BC, IC}\}}+ {\rm MSE}_R$. For solving forward problems, the first term represents a mismatch of the NN output $\mathbf{u}(x,t)$ from boundary and/or initial conditions, i.e., ${\rm MSE}_{\rm BC, IC}$. For solving inverse problems, the first term considers a mismatch of the NN output $\mathbf{u}(x,t)$ from additional data sets, i.e., ${\rm MSE}_u$.

 In general, PINNs contain three steps to solve a physical problem involving PDEs:
\\\indent
Step 1. Define the PINN architecture.
\\\indent
Step 2. Define the loss function ${\rm MSE} = {\rm MSE}_{\{u,{\rm BC, IC}\}} + {\rm MSE}_R$.
\\\indent
Step 3. Train the PINN using an appropriate optimizer, i.e., Adam~\cite{2014Kingma}, AdaGrad~\cite{2011Duchi}.  
\\\noindent
Recently, PINNs and their variants have been successfully applied to solve both forward and inverse problems involving PDEs. Examples include the Navier-Stokes and the KdV equations~\cite{2019Raissi}, stochastic PDEs~\cite{yang2018physics,2018Zhang,zhang2019learning,yang2019adversarial}, and fractional PDEs~\cite{pang2018fpinns}.

In modeling problems involving long-time integration of PDEs, the large number of spatio-temporal degrees of freedom leads to a  large size of data for training the PINN. This will require that PINNs  solve long-time physical problems, which may be computationally prohibitive.
To this end, we propose a {\em parareal physics-informed neural network} (PPINN) to split one long-time problem into many independent short-time problems supervised by an inexpensive/fast coarse-grained (CG) solver, which is inspired by the original parareal algorithm~\cite{2002Maday} and a more recent supervised parallel-in-time algorithm~\cite{2019Blumers}. Because the computational cost of training a DNN increases fast with the size of data set~\cite{2014Livni}, this PPINN framework is able to maximize the benefit of high computational efficiency of training a neural network with small-data sets. More specifically, there is a two-fold benefit from training PINNs with small-data sets rather than working on a large-data set directly, i.e., training of individual PINNs with small-data is much faster, while training of fine PINNs can be readily parallelized on multiple GPU-CPU cores. Consequently, compared to the original PINN approach, the proposed PPINN approach may achieve a significant speed-up for solving long-time physical problems; this will be verified with benchmark tests for one-dimensional and two-dimensional nonlinear problems. This favorable speed-up will depend on a good supervisor that will be represented by a coarse-grained (CG) solver expected to provide reasonable accuracy.

The remainder of this paper is organized as follows. In Section~\ref{sec:method} we describe the details of the PPINN framework and its implementation. In Section~\ref{sec:result} we first present a pedagogical example to demonstrate accuracy and convergence for an one-dimensional time-dependent problem. Subsequently, we apply PPINN to solve a two-dimensional nonlinear time-dependent problem, where we demonstrate the speed-up of PPINN. Finally, we end the paper with a brief summary and discussion in Section~\ref{sec:summary}.

\section{Parareal PINN}\label{sec:method}
\subsection{Methodology}
For a time-dependent problem involving long-time integration of PDEs for $t\in [0,T]$, instead of solving this problem directly in one single time domain, PPINN splits $[0,T]$ into $N$ sub-domains with equal length $\Delta T = T/N$. Then, PPINN employs two propagators, i.e., a serial CG solver represented by $\mathcal{G}(\mathbf{u}_{i}^k)$ in Algorithm~\ref{alg:pPINN}, and $N$ fine PINNs computing in parallel represented by $\mathcal{F}(\mathbf{u}_{i}^k)$ in Algorithm~\ref{alg:pPINN}. Here, $\mathbf{u}_{i}^k$ denotes the approximation to the exact solution at time $t_i$ in the $k$-{th} iteration. Because the CG solver is serial in time and fine PINNs run in parallel, to have the optimal computational efficiency, we encode a simplified PDE (sPDE) into the CG solver as a prediction propagator while the true PDE is encoded in fine PINNs as the correction propagator.
Using this prediction-correction strategy, we expect the PPINN solution to converge to the true solution after a few iterations. 

\begin{figure}[ht!]
  \centering
  \includegraphics[width=0.33\textwidth]{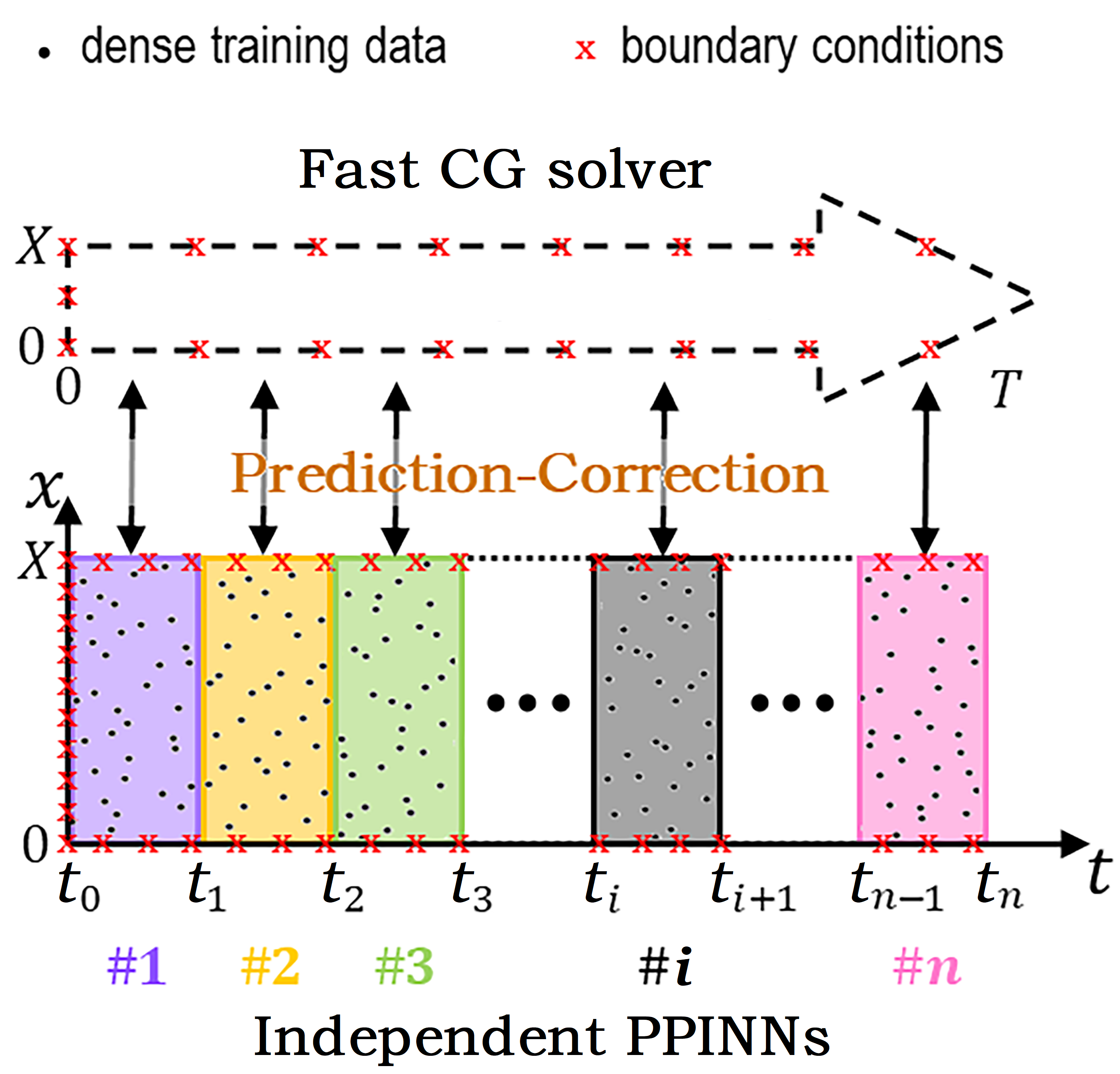}
  \includegraphics[width=0.60\textwidth]{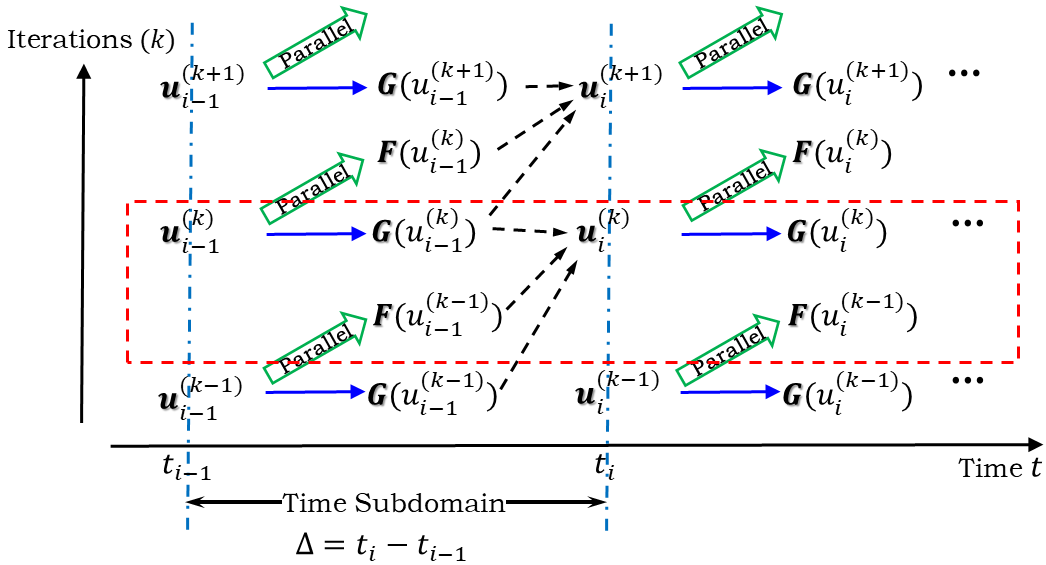}\\
  \caption{Overview of the parareal physics-informed neural network (PPINN) algorithm. Left: Schematic of the PPINN, where a long-time problem (PINN with full-sized data) is split into many independent short-time problems (PINN with small-sized data) guided by a fast coarse-grained (CG) solver. Right: A graphical representation of the parallel-in-time algorithm used in PPINN, in which a cheap serial CG solver is used to make an approximate prediction of the solution $\mathcal{G}(\mathbf{u}_i^k)$, while many fine PINNs are performed in parallel for getting $\mathcal{F}(\mathbf{u}_i^k)$ to correct the solution iteratively. Here, $k$ represents the index of iteration, and $i$ is the index of time subdomain.}
  \label{fig:TP-PINN}
\end{figure}

The details for the PPINN approach are displayed in Algorithm~\ref{alg:pPINN} and Fig.~\ref{fig:TP-PINN}, which are explained step by step in the following: Firstly, we simplify the PDE to be solved by the CG solver. For example, we can replace the nonlinear coefficient in the diffusion equation shown in Sec.~\ref{sec:diffusion} with a constant to remove the variable/multiscale  coefficients. For instance, we can use a CG PINN as the fast CG solver but we can also explore standard fast finite difference solvers. Secondly, the CG PINN is employed to solve the sPDE serially for the entire time-domain to obtain an initial solution. Due to the fact that we can use less residual points as well as smaller neural networks to solve the sPDE rather than the original PDE, the computational cost in the CG PINN can be significantly reduced. Thirdly, we decompose the time-domain into $N$ subdomains. We assume that $\mathbf{u}_i^k$ is known for $t_k \le t_i \leq t_N$ (including $k = 0$, i.e., the initial iteration), which is employed as the initial condition to run the $N$ fine PINNs in parallel. Once the fine solutions at all $t_i$ are obtained, we can compute the discrepancy between the coarse and fine solution at $t_i$ as shown in Step~3(b) in Algorithm~\ref{alg:pPINN}. We then run the CG PINN serially  to update the solution $\mathbf{u}$ for each interface between two adjacent subdomains, i.e., $\mathbf{u}_{i+1}^{k+1}$ (Prediction and Refinement in Algorithm~\ref{alg:pPINN}). Step 3 in Algorithm~\ref{alg:pPINN} is performed iteratively until the following criterion is satisfied 

\begin{equation}
    E = \frac{\sqrt{\sum^{N-1}_{i=0} ||\mathbf{u}_{i}^{k+1} - \mathbf{u}_{i}^{k}||^2}}{\sqrt{\sum^{N-1}_{i=0} ||\mathbf{u}_{i}^{k+1}||^2}} < E_{tol},
\end{equation}
where $E_{tol}$ is a user-defined tolerance, which is set as $1\%$ in the present study.  

\begin{algorithm}[h!]
\caption{PPINN algorithm}
\begin{algorithmic}[1]
\State Model reduction: Specify a simplified PDE for the fast CG solver to supervise the fine PINNs.
\State Initialization:

Solve the sPDE with the CG solver for the initial iteration: $\mathbf{u}_{i+1}^0 = \mathcal{G}(\mathbf{u}_{i}^0)$ for all $0\le t_i \le t_N$.

\State Assume $\{\mathbf{u}_{i}^k\}$ is known for $t_k \leq t_i\leq t_N$ and $k\geq 0$:

Correction:

\hspace{\algorithmicindent} (a) Advance with fine PINNs in parallel:  $\mathcal{F}(\mathbf{u}_{i}^k)$ for all $t_k \leq t_i \leq t_{N-1}$.

\hspace{\algorithmicindent} (b) Correction:  $\delta_{i}^k = \mathcal{F}(\mathbf{u}_{i}^k) - \mathcal{G}(\mathbf{u}_{i}^k)$ for all $t_k \le t_i \le t_{N-1}$.

Prediction:

\hspace{\algorithmicindent} Advance with the fast CG solver in serial:  $\mathcal{G}(\mathbf{u}_{i}^{k+1})$ for all $t_{k+1} \le t_i \le t_{N-1}$.

Refinement:

\hspace{\algorithmicindent} Combine the correction and prediction terms: $\mathbf{u}_{i+1}^{k+1} = \mathcal{G}(\mathbf{u}_{i}^{k+1}) + \mathcal{F}(\mathbf{u}_{i}^k) - \mathcal{G}(\mathbf{u}_{i}^k)$

\State Repeat \textit{Step 3} to compute $\mathbf{u}_{i}^{k+2}$ for all $1\le t_i \le t_N$ until a convergence criterion is met.

\end{algorithmic}
\label{alg:pPINN}
\end{algorithm}

\subsection{Speed-up analysis}
\label{sec:speedup}
The walltime for the PPINN can be expressed as $T_{PPINN} = T^0_c + \sum^{K}_{k = 1}( T^k_c + T^k_f )$, where $K$ is the total number of iterations, $T^0_c$ represents the walltime taken by the CG solver for the initialization, while $T^k_c$ and $T^k_f$ denote the walltimes taken by the coarse and fine propagators at $k$-{th} iteration, respectively. Let $\tau^k_c$ and $\tau^k_f$ be the walltimes used by CG solver and fine PINN for one subdomain at $k$-{th} iteration, $T^k_c$ and $T^k_f$ can be expressed as
\begin{align}
    T^k_c = N \cdot \tau^k_c, \quad T^k_f = \tau^k_f,
\end{align}
where $N$ is the number of subdomains. Therefore, the walltime for PPINN is
\begin{equation}
    T_{PPINN} = T^0_c + \sum^{K}_{k = 1}\left( N \cdot \tau^k_c + \tau^k_f \right) = N \cdot \tau^0_c + \sum^{K}_{k = 1}\left( N \cdot \tau^k_c + \tau^k_f \right).
\end{equation}
Furthermore, the walltime for the fine PINN to solve the same PDE in serial is expressed as $T_{PINN} = N \cdot T^1_f$. To this end, we can obtain the speed-up ratio of the PPINN as
\begin{align}
    S = \frac{T_{PINN}}{T_{PPINN}} = \frac{N \cdot T^1_f} {N \cdot \tau^0_c + \sum^{K}_{k = 1}\left( N \cdot \tau^k_c + \tau^k_f \right)},
\end{align}
which can be rewritten as 
\begin{equation}
    S = \frac{N \cdot \tau^1_f}{ N \cdot \tau^0_c + \tau^1_f + N \cdot K \cdot \tau^k_c + (K - 1) \cdot \tau^{k}_f}.
\end{equation}
In addition, considering that the solution in each subdomain for two adjacent iterations ($k \geq 2$) does not change dramatically, the training process converges much faster for $k \geq 2$ than that of $k = 1$ in each fine PINN, i.e., $\tau^k_f \ll \tau^1_f$. Therefore, the lower bound for $S$ can be expressed as
\begin{equation}
    S_{min} = \frac{N \cdot \tau^1_f}{ N \cdot \tau^0_c +  N \cdot K \cdot \tau^k_c + K \cdot \tau^{1}_f}.
\end{equation}
This shows that $S$ increases with $N$ if $\tau^0_c \ll \tau^1_f$, suggesting that the more subdomains we employ, the larger the speed-up ratio for the PPINN.

\section{Results}\label{sec:result}
We first show two simple examples for a deterministic and a stochastic ordinary differential equation (ODE) to explain the method in detail. We
then present results for the Burgers equation and a two-dimensional diffusion-reaction equation.
\subsection{Pedagogical examples}
\label{sec:ode}
We first present two pedagogical examples, i.e., a deterministic and a stochastic ODE, to demonstrate the procedure of using PPINN to solve simple time-dependent problems; the main steps and key features of PPINN can thus be easily understood. 

\subsubsection{Deterministic ODE}
\label{sec:ode1}
The deterministic ODE considered here  reads as
\begin{equation}\label{eq:ODE}
 \frac{du}{dt} = a + \omega \cos(\omega t),
\end{equation}
where the two parameters are $a=1$ and $\omega = \pi/2$. Given an initial condition $u(0)=0$, the exact solution of this ODE is $u(t) = t + \sin(\pi t/2)$. The length of time domain we are interested is $T = 10$.

In the PPINN approach, we decompose the time-domain $t\in [0, 10]$ into $10$ subdomains, with length $\Delta t = 1$ for each subdomain. We use a simplified ODE (sODE) $du/dt = a$ with IC $u(0)=0$ for the CG solver, and use the exact ODE $du/dt = a + \omega\cos(\omega t)$ with IC $u(0)=0$ for the ten fine PINNs. In particular, we first use a CG PINN to act as the fast solver. Let $[N_I]+[N_H]\times D+[D_O]$ represent the architecture of a DNN, where $N_I$ is the width of the input layer, $N_H$ is the width of the hidden layer, $D$ is the depth of the hidden layers and $N_O$ is the width of the output layer. The CG PINN is constructed as a $[1]+[4]\times 2+[1]$ DNN encoded with the sODE $du/dt = a$ and IC $u(0)=0$, while each fine PINN is constructed as a $[1]+[16]\times 3 + [1]$ DNN encoded with the exact ODE $du/dt = a + \omega\cos(\omega t)$ and IC $u(0)=0$. 

\begin{figure}[b!]
  \centering
  \subfigure[]{\label{fig:ODEa}
  \includegraphics[width=0.45\textwidth]{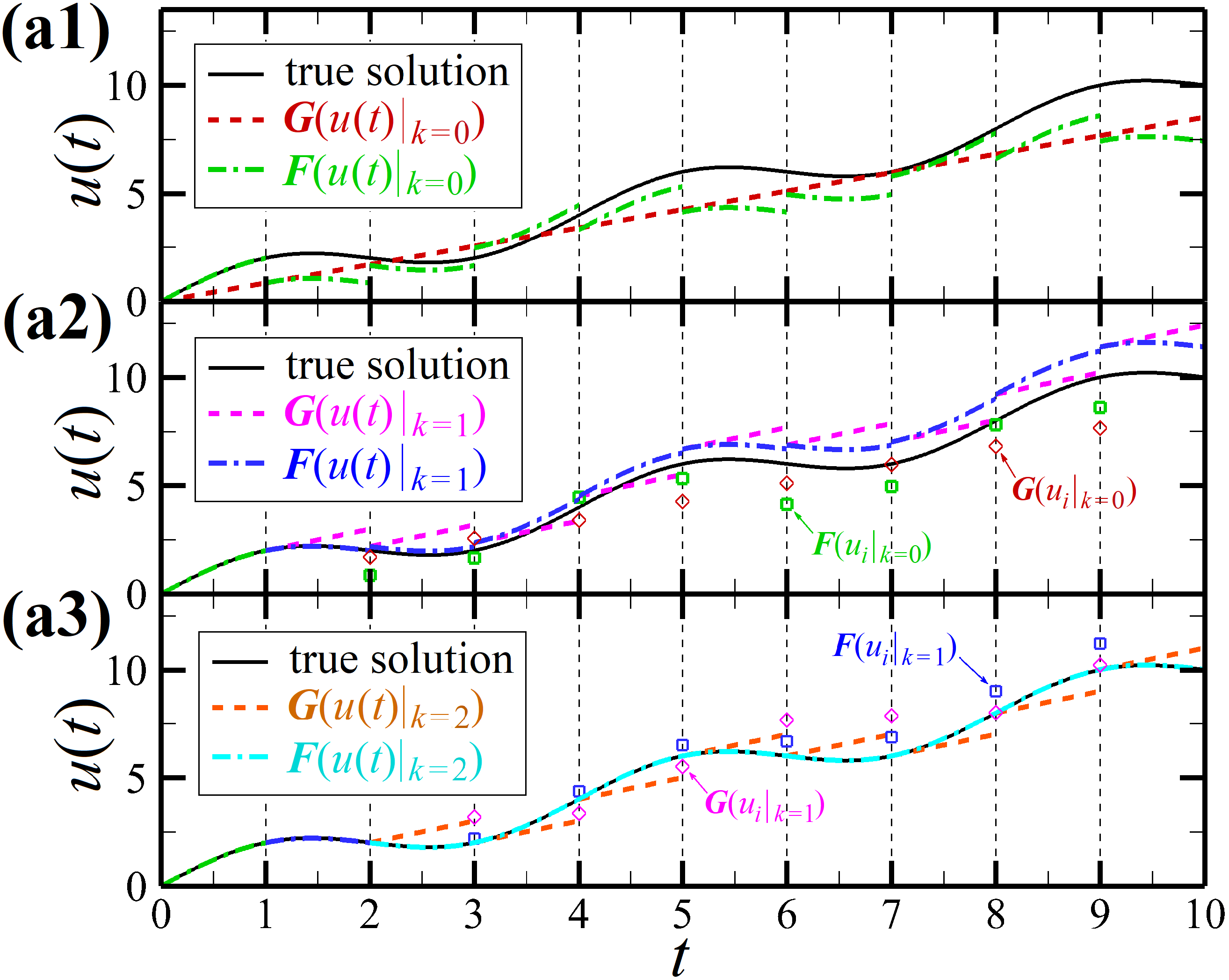}}
  \subfigure[]{\label{fig:ODEb}
  \includegraphics[width=0.45\textwidth]{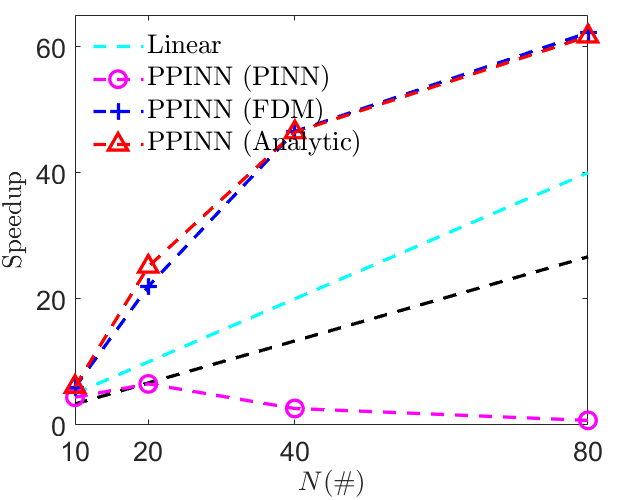}}
  \caption{ \label{fig:ODE}
  A pedagogical example to illustrate the procedure of using PPINN  to solve a deterministic ODE. 
  Here a PINN is also employed for the CG solver.\ 
  (a) The solutions of CG PINN $\mathcal{G}(u(t)|_k)$ and fine PINNs $\mathcal{F}(u(t)|_k)$ in different iterations are plotted for (a1) $k = 0$, (a2) $k=1$ and (a3) $k=2$ to solve the ODE $du/dt = 1 + \pi/2\cdot\cos(\pi t/2)$. (b) Speed-ups for the PPINN using the different coarse solvers, i.e. PINN (magenta dashed line with circle), FDM (blue dashed line with plus), and analytic solution (red dashed line with triangle). $N$ denotes the number of the subdomains. The linear speed-up ratio is calculated as $N/K$.  Black dashed line: $K = 3$, Cyan dashed line: $K = 2$.
 }
\end{figure}

In the first iteration, we train the CG PINN for $5,000$ epochs using the Adam algorithm with a learning rate of $\alpha = 0.001$, and obtain a rough prediction of the solution $\mathcal{G}(u(t))|_{k=0}$, as shown in Fig.~\ref{fig:ODE}(a1). Let $t_i = i$ for $i = 0,1,\dddot\, , 9$ be the starting time of each fine PINN in the chunk $i+1$. To compute the loss function of the CG PINN, $20$ residual points uniformly distributed in each subdomain $[t_i,t_{i+1}]$ are used to compute the residual ${\rm MSE}_R$. Subsequently, with the CG PINN solutions, the IC of each fine PINN is given by the predicted values $\mathcal{G}(u(t))|_{k=0}$ at $t = t_i$ except the first subdomain. Similarly to training the CG PINN, we train ten individual fine PINNs in parallel using the Adam algorithm with a learning rate of $\alpha = 0.001$. We accept the PINN solution if the loss function of each fine PINN $loss = {\rm MSE}_R + {\rm MSE}_{\rm IC} < 10^{-6}$, wherein $101$ residual points uniformly distributed in $[t_i, t_{i+1}]$ are used to compute the residual ${\rm MSE}_R$. We can observe in Fig.~\ref{fig:ODE}(a1) that the fine PINN solutions correctly capture the shape of the exact solution, but their magnitudes significantly deviate from the exact solution because of the poor quality of the IC given by $\mathcal{G}(u(t))|_{k=0}$.

The deviation of PPINN solution from the exact solution is quantified by the $l_2$ relative error norm
\begin{equation}\label{eq:l2}
 l_2 = \frac{\sum_p(u^k_p - \hat{u}_p)^2}{\sum_p \hat{u}_p^2},
\end{equation}
where $p$ denotes the index of residual points. The PPINN solution after the first iteration presents obvious deviations, as shown in Fig.~\ref{fig:ODE}(a1). In the second iteration, given the solutions of $\mathcal{G}(u_i)|_{k=0}$ and $\mathcal{F}(u_i)|_{k=0}$, we set the IC for CG PINN as $G$. Then, we train the CG PINN again for $5,000$ epochs using a learning rate $\alpha = 0.001$ and obtain the solutions $\mathcal{G}(u(t))|_{k=1}$ for the chunks $2$ to $10$, as shown in Fig.~\ref{fig:ODE}(a2). The IC for fine PINN of the second chunk is given by $\tilde{u}(t = t_1) = F(u_1)_{k=1}$, and the ICs of fine PINNs of the chunks $3$ to $10$ are given by a combination of $\mathcal{G}(u(t))|_{k=0}$, $\mathcal{F}(u(t))|_{k=0}$ and $\mathcal{G}(u(t))|_{k=1}$ at $t = t_{i=2,3\dddot\, , 9}$, i.e., $\tilde{u}(t = t_i) = \mathcal{G}(u_i)|_{k=1} - \left[\mathcal{F}(u_i)|_{k=0} - \mathcal{G}(u_i)|_{k=0}\right]$. With these ICs, we train the nine individual fine PINNs (chunks $2$ to $10$) in parallel using the Adam algorithm with a learning rate of $\alpha = 0.001$ until the loss function for each fine PINN drops below $10^{-6}$. We find that the fine PINNs in the second iteration converge much faster than in the first iteration, with an average ecochs of $4836$ compared to $11524$ in the first iteration. The accelerated training process benefits from the results of training performed in the previous iteration. The PPINN solutions for each chunk after the second iteration are presented in Fig.~\ref{fig:ODE}(a2), showing a significant improvement of the solution with a relative error $l_2 = 9.97 \%$. Using the same method, we perform a third iteration, and the PPINN solutions converge to the exact solution with a relative error $l_2 = 0.08 \%$, as shown in Fig.~\ref{fig:ODE}(a3).

We proceed to investigate the computational efficiency of the PPINN. As mentioned in Sec.~\ref{sec:speedup}, we may obtain a speed-up ratio, which grows linearly with the number of subdomains if the coarse solver is efficient enough. We first test the speed-ups for the PPINN using a PINN as the coarse solver. Here we test four different subdomain decompositions, i.e. 10, 20, 40, and 80. For the coarse solver, we assign one PINN (CG PINN, [1] + [4] $\times$ 2 + [1]) for each subdomain, and 10 randomly sampled residual points are used in each subdomain.  Furthermore, we run all the CG PINNs serially using one CPU (Intel Xeon E5-2670). For the fine solver, we again employ one PINN (fine PINN) for each subdomain to solve Eq.~\eqref{eq:ODE}, and  each subdomain is assigned to one CPU (Intel Xeon E5-2670). The total number for the residual points is 400,000, which are randomly sampled in the entire time domain and will be uniformly distributed in each subdomain. Meanwhile, the architecture for the fine PINN in each subdomain is $[1] + [20] \times 2 + [1]$ for the first two cases (i.e., 10 and 20 subdomains), which is then set as $[1] + [10] \times 2 + [1]$ for the last two cases (i.e., 40 and 80 subdomains). The speed-ups are displayed in Fig. \ref{fig:ODEb} (magenta dashed line), where we observe that the speed-up first increases with $N$ as $N \le 20$, then it decreases with the increasing $N$. We further look into the details of the computational time for this case. As shown in Table \ref{tab:speedup_ode_pinn}, we found that the computational time taken by the CG PINNs increases with the number of the subdomains. In particular, more than 90 $\%$ of the computational time is taken by the coarse solver as $N \ge 40$, suggesting the inefficiency of the CG PINN. To obtain satisfactory speed-ups using the PPINN, a more efficient coarse solver should be utilized.

\begin{table}[htbp]
\centering
 \begin{tabular}{c|ccccc}
  \hline \hline
  Subdomains ($\#$) & Iterations ($\#$) & $\mathcal{N}_{CG}$ ($\#$) & $T_{CG} (s)$ & $T_{total}$ (s) & $S$\\ \hline
 1    &  - & - & - & 1,793.0 & - \\
 10    &  2 & 10 & 62.4 & 407.4 & 4.4\\
 20    &  2  & 5 & 189.2 & 275.1 & 6.5 \\
 40    &  3 & 4 & 617.8 & 685.2 & 2.6 \\
 80    &  3 & 4 & 2,424.6 & 2,453.0 & 0.73\\
  \hline \hline
 \end{tabular}

  \caption{\label{tab:speedup_ode_pinn}
  Speed-ups for the PPINN using a PINN as coarse solver to solve Eq.~\eqref{eq:ODE}. $\mathcal{N}_{CG}$ is the number of residual points used for the CG PINN in each subdomain, $T_{CG}$ represents the computational time taken by the coarse solver, and $T_{total}$ denotes the total computational time taken by the PPINN.}
\end{table}

Considering that the simplified ODE can be solved analytically, we can thus directly use the analytic solution for the coarse solver which has no cost. In addition,  all parameters in the fine PINN are kept the same as the previously used ones. For validation, we again use 10 subdomains in the PPINN to solve  Eq. \eqref{eq:ODE}. The $l_2$ relative error between the predicted and analytic solutions is $1.252 \times 10^{-5}$ after two iterations, which confirms the accuracy of the PPINN. In addition, the speed-ups for the four cases are displayed in Fig.~\ref{fig:ODEb} (red dashed line with triangle). It is interesting to find that the PPINN can achieve a superlinear speed-up here. We further present the computational time at each iteration in Table \ref{tab:speedup_ode_fdm}. We see that the computational time taken by the coarse solver is now negligible compared to the total computational time. Since the fine PINN converges faster after the first iteration, we can thus obtain a superlinear speed-up for the PPINN.

Instead of the analytic solution, we can also use other efficient numerical methods to serve as the coarse solver. For demonstration purpose, we then present the results using the finite difference method (FDM) as the coarse solver. The entire domain is discretized into 1,000 uniform elements, which are then  uniformly distributed to each subdomain. Similarly, we also use 10 subdomains for validation. It also takes 2 iterations to converge, and the $l_2$ relative error is $1.257 \times 10^{-5}$. In addition, we can also obtain a superlinear speed-up which is quite similar to the case using the analytic solution due to the efficiency of the FDM (Fig. \ref{fig:ODEb} and Table \ref{tab:speedup_ode_fdm}). 

\begin{table}[htbp]
\centering
 \begin{tabular}{c|cccccc}
  \hline \hline
  &Subdomains ($\#$) & Iterations ($\#$) & $\mathcal{N}_{CG}$ ($\#$) & $T_{CG} (s)$ & $T_{total}$ (s) & $S$\\ \hline
 \multirow{5}{*}{PPINN (Analytic)} 
 &1&  - & - & - & 1,793.0 & - \\
&10  &  2 & - & $< 0.05$ & 295.8 & 6.1\\
&20  &  2  & - & $< 0.05$ & 71.4 & 25.1 \\
&40  &  2 & - & $< 0.05$ & 38.6 & 46.4 \\
&80 &  2 & - & $< 0.05$ & 29.0 & 61.6 \\ \hline
  \multirow{5}{*}{PPINN (FDM)} 
  &1&  - & - & - & 1,793.0 & - \\
&10  &  2 & 100 & $< 0.05$ & 298.7 & 6.0 \\
&20  &  2  & 50 & $< 0.05$ & 81.6 & 22.0 \\
&40  &  2 & 40 & $< 0.05$ & 38.5 & 46.5 \\
&80 &  2 & 12 & $< 0.05$ & 28.8 & 62.2\\                         
  \hline \hline
 \end{tabular}
  \caption{\label{tab:speedup_ode_fdm} Walltimes for using the PPINN with different coarse solvers to solve Eq. ~\eqref{eq:ODE}. $\mathcal{N}_{CG}$ is the number of elements used for the FDM in each subdomain, $T_{CG}$ represents the computational time taken by the coarse solver, and $T_{total}$ denotes the total computational time taken by the PPINN.}
\end{table}

\subsubsection{Stochastic ODE}
\label{sec:ode2}
In addition to the deterministic ODE, we can also apply this idea to solve stochastic ODEs. Here, we consider the following stochastic ODE 
\begin{align}\label{eq:sode}
    \frac{du}{dt} = \beta \left[ -u + \beta \sin(\frac{\pi}{2} t) + \frac{\pi}{2} \cos(\frac{\pi}{2} t) \right], t \in [0, T],
\end{align}
where $T = 10$, $\beta = \beta_0 + \epsilon$, $\beta_0 = 0.1$, and $\epsilon$ is drawn from a normal distribution with zero mean and 0.05 standard deviation. In addition, the initial condition for Eq. \eqref{eq:sode} is $u(0) = 1$. 

In the present PPINN, we employ a deterministic ODE for the coarse solver as 
\begin{align}\label{eq:sode_s}
    \frac{du}{dt} = - \beta_0 u, t \in [t_0, T].  
\end{align}
Given the initial condition $u(t_0) = u_0$, we can obtain the analytic solution for Eq.~\eqref{eq:sode_s} as $u = u_0 \exp(- \beta_0 (t - t_0))$. For the fine solver, we draw 100 samples for the $\beta$ using the quasi-Monte Carlo method, which are then solved by the fine PINNs. Similarly, we utilize three different methods for the coarse solver, i.e. the PINN, FDM, and analytic solution.

\begin{figure}[t!]
  \centering
  \subfigure[]{\label{fig:sode_speedupa}
  \includegraphics[width=0.45\textwidth]{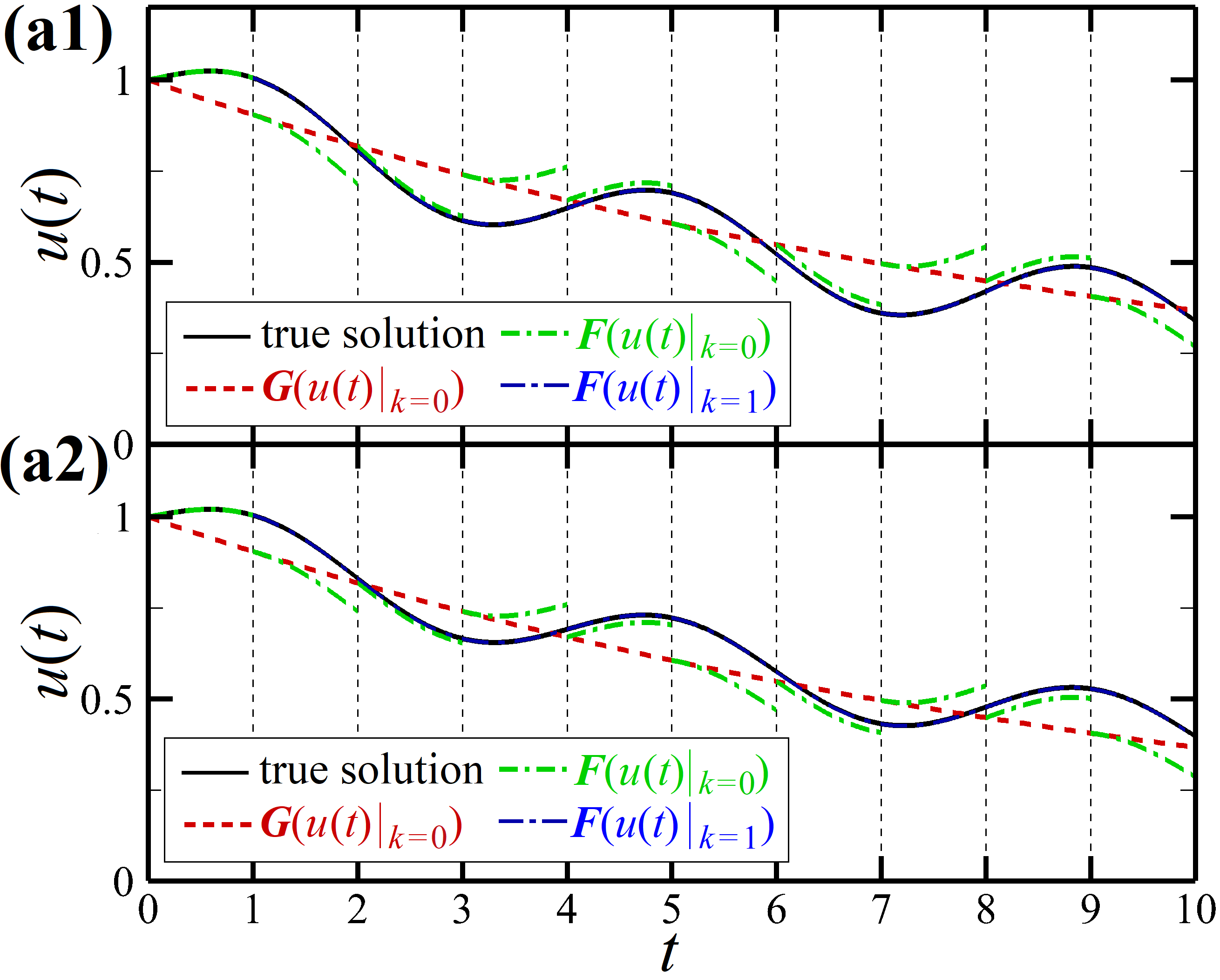}}
  \subfigure[]{\label{fig:sode_speedupb}
  \includegraphics[width=0.45\textwidth]{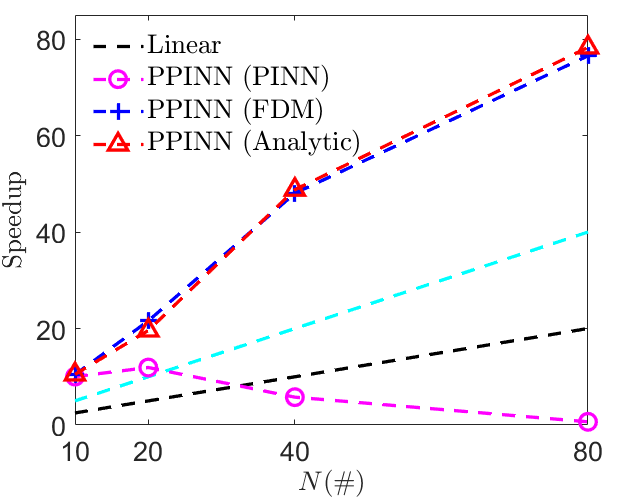}}
  \caption{A pedagogical example for using the PPINN to solve stochastic ODE.  (a) The solutions of CG solver $\mathcal{G}(u(t)|_k)$ and fine PINNs $\mathcal{F}(u(t)|_k)$ in different iterations are plotted for (a1) $\beta = 0.108$, and (a2) $\beta = 0.090$. The reference solution is $u(t) = \exp(- \beta t) + \beta \sin(\frac{\pi}{2}t)$ for each $\beta$. (b) Speed-ups for the PPINN using different coarse solvers, i.e. the PINN (magenta dashed line with circle), the FDM (blue dashed line with plus), and analytic solution (red dashed line with triangle). The linear speed-up ratio is calculated as $N/K$. Black dashed line: $K = 4$, Cyan dashed line: $K =2$. For the CG PINN, we use 1,000 randomly sampled residual points in the whole time domain, which are uniformly distributed to the subdomains. The architecture of the PINN for each subdomain is the same, i.e. $[1] + [10] \times 2 + [1]$. For the fine PINN, the architecture is $[1] + [20] \times 2 + [1]$ for the cases with 10 and 20 subdomains, and it is $[1] + [10] \times 2 + [1]$ for the cases with 40 and 80 subdomains.}
  \label{fig:sode_speedup}
\end{figure}

For validation purposes, we first decompose the time-domain into 10 uniform subdomains. For the FDM, we discretize the whole domain into 1,000 uniform elements. For the fine PINNs, we employ 400,000 randomly sampled residual points for the entire time domain, which are uniformly distributed to all the subdomains.  We employ one fine PINN for each subdomain to solve the ODE, which has an architecture of $[1] + [20] \times 2 + [1]$. Finally, the simplified ODE in each subdomain for the coarse solver is solved serially, while the exact ODE in each subdomain for the fine solver is solved in parallel. We illustrate the comparison between the predicted and exact solutions at two representative $\beta$, i.e. $\beta = 0.108$ and $0.090$ in Fig.~\ref{fig:sode_speedupa}. We see that the predicted solutions converge to the exact ones after two iterations, which confirms the effectiveness of the PPINN for solving stochastic ODEs. The solutions from the PPINN with the other two different coarse solvers (i.e., the PINN and analytic solution) also agree well with the reference solutions, but they are not presented here. Furthermore, we also present the computational efficiency for the PPINN using four different numbers of subdomains, i.e. $10, 20, 40$ and $80$ in Fig.~\ref{fig:sode_speedupb}. It is clear that the PPINN can still achieve a superlinear speed-up if we use an efficient coarse solver such as FDM or analytic solution. The speed-up for the PPINN with the PINN as coarse solver is similar to the results in Sec.~\ref{sec:ode1}, i.e., the speed-up ratio first slightly increases with the number of  subdomains as $N \le 20$, then it decreases with the increasing $N$. 
Finally, the speed-ups for the PPINN with FDM coarse solver are almost the same as the PPINN with analytic solution coarse solver, which are similar as the results in Sec.~\ref{sec:ode1} and will not be discussed in detail here.

In summary, the PPINN can work for both deterministic and stochastic ODEs. In addition, using the PINN as the coarse solver can guarantee the accuracy for solving the ODE, but the speed-up may decrease with the number of subdomains due to the inefficiency of the PINN. We can achieve  both high accuracy and good speed-up if we select more efficient coarse solvers, such as an analytic  solution, a finite difference method, and so on.

\subsection{Burgers equation}
\label{sec:burgers}
We now consider the viscous Burgers equation
\begin{equation}\label{eq:burgers}
 \frac{\partial u}{\partial t} + u\frac{\partial u}{\partial x} = \nu \frac{\partial^2 u}{\partial x^2},
\end{equation}
which is a mathematical model for the viscous flow, gas dynamics, and traffic flow, with $u$ denoting the speed of the fluid, $\nu$ the kinematic viscosity, $x$ the spatial coordinate and $t$ the time. Given an initial condition $u(x,0)=-\sin(\pi x)$ in a domain $x\in[-1,1]$, and the boundary condition $u(\pm 1, t) = 0$ for $t\in [0,1]$, the PDE we would like to solve is Eq.~\eqref{eq:burgers} with a viscosity of $\nu = 0.03/\pi$.

In the PPINN, the temporal domain $t \in [0, 1]$ is decomposed into 10 uniform subdomains. Each subdomain has a time length $\Delta t = 0.1$. The simplified PDE for the CG PINN is also a Burgers equation, which uses the same initial and boundary conditions but with a larger viscosity $\nu_c = 0.05/\pi$. It is well known that the Burgers equation with a small viscosity will develop steep gradient in its solution as time evolves.  The increased viscosity will lead to a smoother solution, which can be captured using much less computational cost. Here, we use the same NN for the CG and fine PINNs for each subdomain, i.e., 3 hidden layers with 20 neurons per layer. The learning rates are also set to be the same, i.e., $10^{-3}$. Instead of using one optimizer in the last case, here we use two different optimizations, i.e., we first train the PINNs using the Adam optimizer (first-order convergence rate) until the loss is less than $10^{-3}$, then we proceed to employ the L-BFGS-B method to further train the NNs. The L-BFGS is a quasi-Newtonian approach which has second-order convergence rate and can enhance the convergence of the training \cite{2019Raissi}.

\begin{figure}[t!]
  \centering
  \subfigure[]{\label{fig:Burgersa}
  \includegraphics[width=0.45\textwidth]{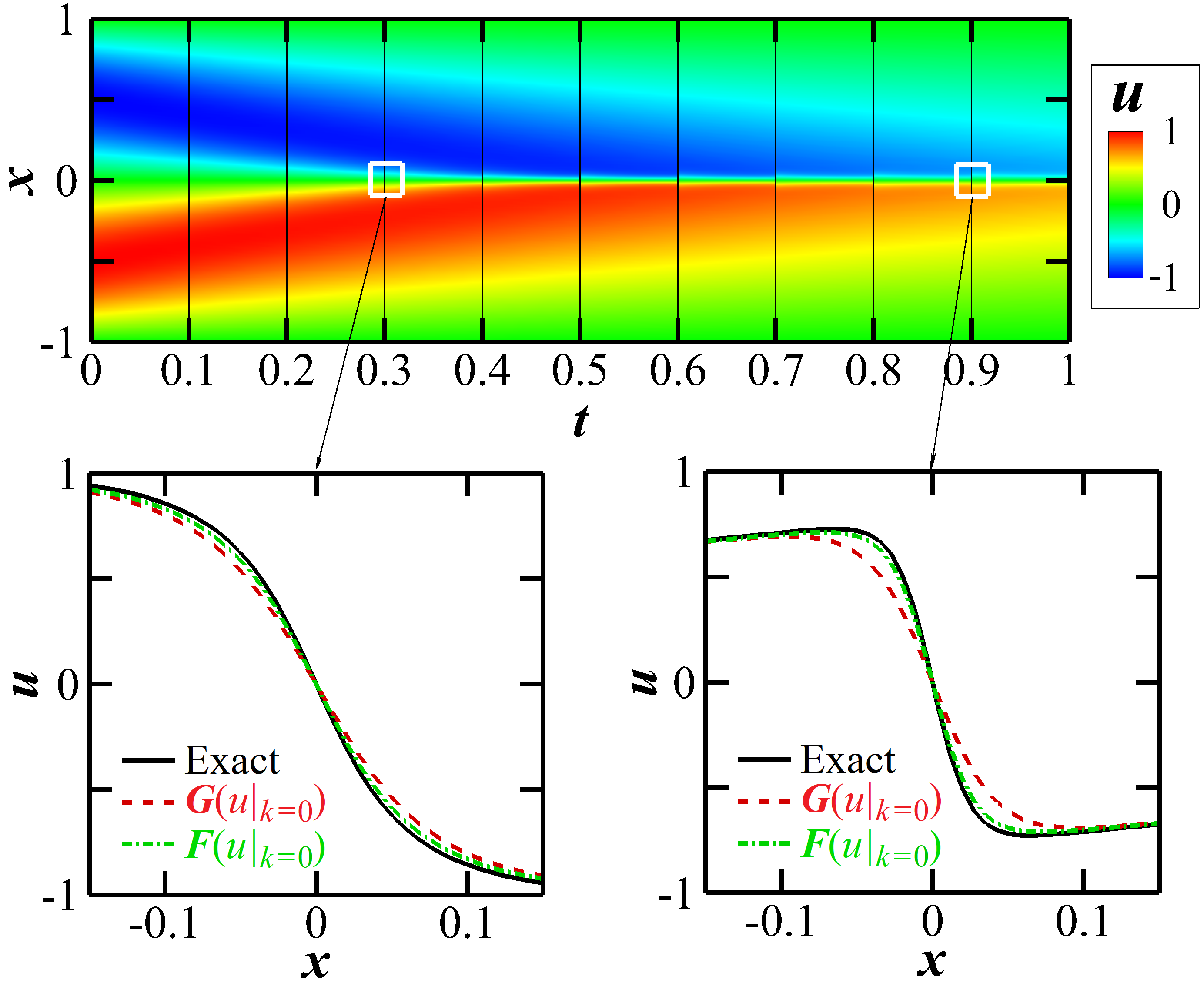}}
  \subfigure[]{\label{fig:Burgersb}
  \includegraphics[width=0.45\textwidth]{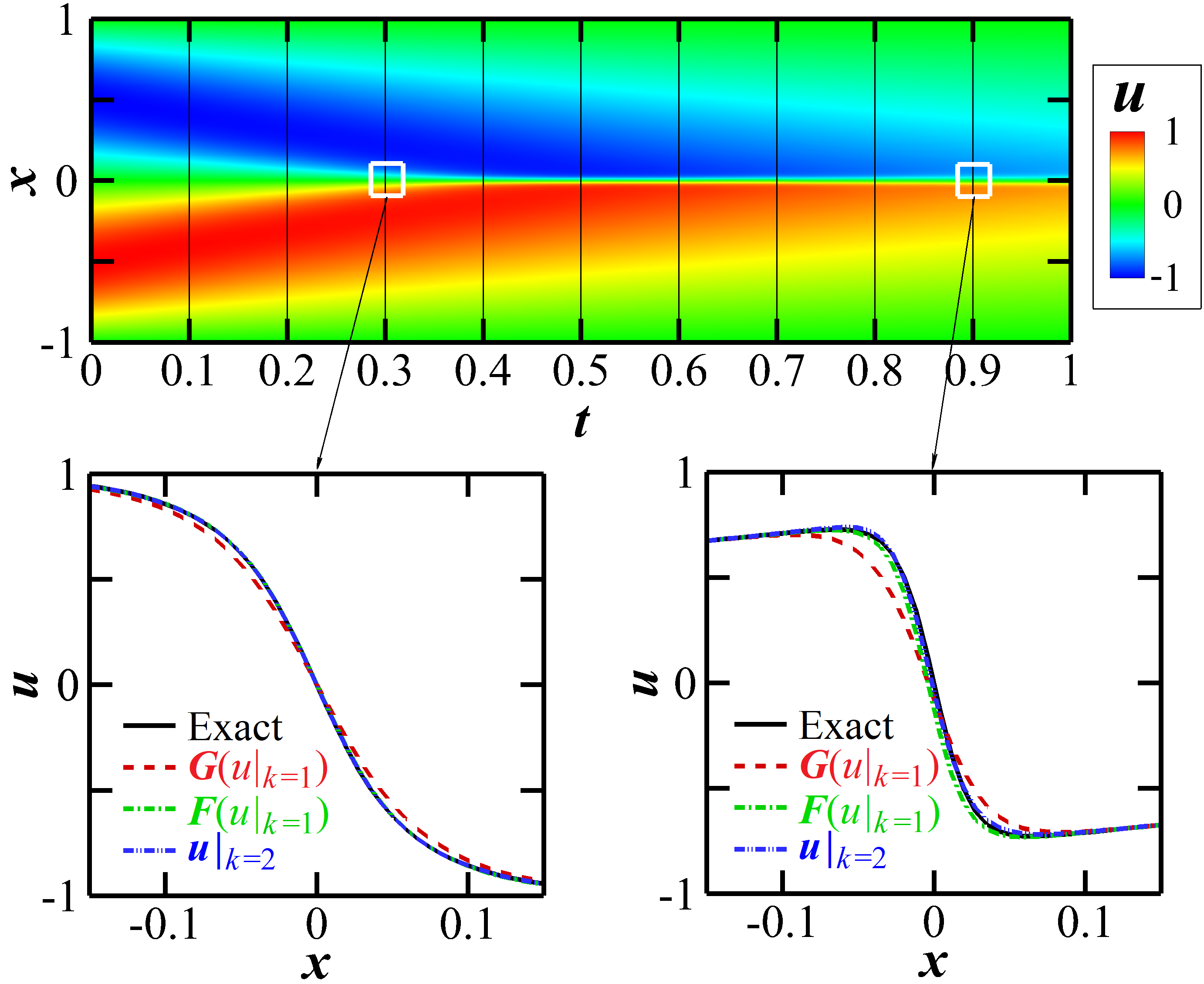}}
  \caption{Example of using the PPINN for solving the Burgers equation.\ $\mathcal{G}(u(t)|_k)$ represents the rough prediction generated by the CG PINN in the $(k+1)$-th iteration, while $\mathcal{F}(u(t)|_k)$ is the solution corrected in parallel by the fine PINNs. (a) Predictions after the first iteration $(k=0)$ at $t = 0.3$ and 0.9), (b) Predictions after the second iteration ($k=1$) at $t = 0.3$ and 0.9.}
  \label{fig:Burgers}
\end{figure}

For the CG PINN in each subdomain, we use 300 randomly sampled residual points to compute the $\mbox{MSE}_{R}$, while $1,500$ random residual points are employed in the fine PINN for each subdomain. In addition, 100 uniformly distributed points are employed to compute the $\mbox{MSE}_{IC}$ is  for each subdomain, and 10 randomly sampled points are used to compute the $\mbox{MSE}_{BC}$ in both the CG and fine PINNs. For this particular case, it takes only 2 iterations to meet the convergence criterion, i.e., $E_{tol} = 1 \%$. The distributions of the $u$ at each iteration are plotted in Fig.~\ref{fig:Burgers}. As shown in Fig.~\ref{fig:Burgersa}, the solution from the fine PINNs ($\mathcal{F}(u|_{k=0})$) is different from the exact one, but the discrepancy is observed to be small. It is also observed that the solution from the CG PINNs is smoother than that from the fine PINNs especially for solutions at large times, e.g., $t = 0.9$. In addition, the velocity at the interface between two adjacent subdomains is not continuous due to the inaccurate initial condition for each subdomain at the first iteration. At the second iteration (Fig.~\ref{fig:Burgersb}), the solution from the Refinement step i.e. $u|_{k=2}$  shows little difference from the exact one, which confirms the effectiveness of the PPINN. Moreover, the discontinuity between two adjacent subdomains is significantly reduced. Finally, it is also interesting to find that the number of the training steps for each subdomain at the first iteration is from ten to hundred thousand, but they decrease to a few hundred at the second iteration.  Similar results are also reported and analyzed in Sec.~\ref{sec:ode}, which will not be presented here again.

To test the flexibility of the PPINN, we further employ two much larger viscosities in the coarse solver, i.e. $\nu_c = 0.09/\pi$ and $0.12/\pi$, which are $3 \times$ and $4 \times $ the exact viscosity, respectively. All the parameters (e.g., the architecture of the PINN, the number of residual points, etc.) in these two cases are kept the same as the case with $\nu_c = 0.05/\pi$. As shown in Table \ref{tab:burgers}, it is interesting to find that the computational errors for these three cases are comparable, but the number of iterations increases with the viscosity employed in the coarse solver. Hence, in order to obtain an optimum strategy in selecting the CG solver we have to consider the trade-off between the accuracy that the CG solver can obtain and the number of iterations required for convergence of the PPINN.

\begin{table}[htbp]
\centering
 \begin{tabular}{c|ccc}
  \hline \hline
  $\nu_c$  & $0.05/\pi$  & $0.09/\pi$ & $0.12/\pi$\\ \hline
 Iterations ($\#$) & 2 & 3 & 4 \\ \hline
 $l_2 ~(\%)$ & 0.13 & 0.19 & 0.22 \\ \hline
 \hline \hline
 \end{tabular}

  \caption{\label{tab:burgers}
  The PPINN for solving the Burgers equation with different viscosities in the coarse solver. $\nu_c$ represents the viscosity employed in the coarse solver.}
\end{table}

\subsection{Diffusion-reaction equation}
\label{sec:diffusion}
We consider the following two-dimensional diffusion-reaction equation 
\begin{align}\label{reaction}
    \partial_t C = \nabla \cdot (D \nabla C) &+ 2 A\sin(\frac{\pi x}{l})\sin(\frac{\pi y}{l}), ~x,~y \in [0, l], ~ t \in [0, T], \\
    C(x, y, t = 0) &= 0, \\
    C(x = 0, y = 0, t) &= C(x = 0, y = l, t) = C(x = l, y = 0, t) = C(x = l, y = l, t) = 0,
\end{align}
where $l = 1$ is the length of the computational domain, $C$ is the solute concentration, $D$ is the diffusion coefficient, and $A$ is a constant. Here $D$ depends on $C$ as follows:
\begin{align}
    D = D_0 \exp(RC),
\end{align}
where $D_0 = 1 \times 10^{-2}$.

We first set $T = 1$, $A = 0.5511$, and $R = 1$ to validate the proposed algorithm for the two-dimensional case. In the PPINN, we use the PINNs for both the coarse and fine solver. The time domain is divided into 10 uniform subdomains with $\Delta t = 0.1$ for each subdomain. For the coarse solver, the following simplified PDE $\partial_t C = D_0 \nabla^2 C  + 2 A\sin({\pi x}/{l})\sin({\pi y}/{l})$ is solved with the same initial and boundary conditions described in Eq.~\eqref{reaction}. The diffusion coefficient for the coarse solver is about 1.7 times smaller than the maximum $D$ in the fine solver.  The architecture of the NNs employed in both the coarse and fine solvers for each subdomain is kept the same, i.e., 2 hidden layers with 20 neurons per layer. The employed learning rate as well as the optimization method are the same as those applied in Sec.~\ref{sec:burgers}.   

\begin{figure}[h!]
    \centering
    \includegraphics[width=0.8\textwidth]{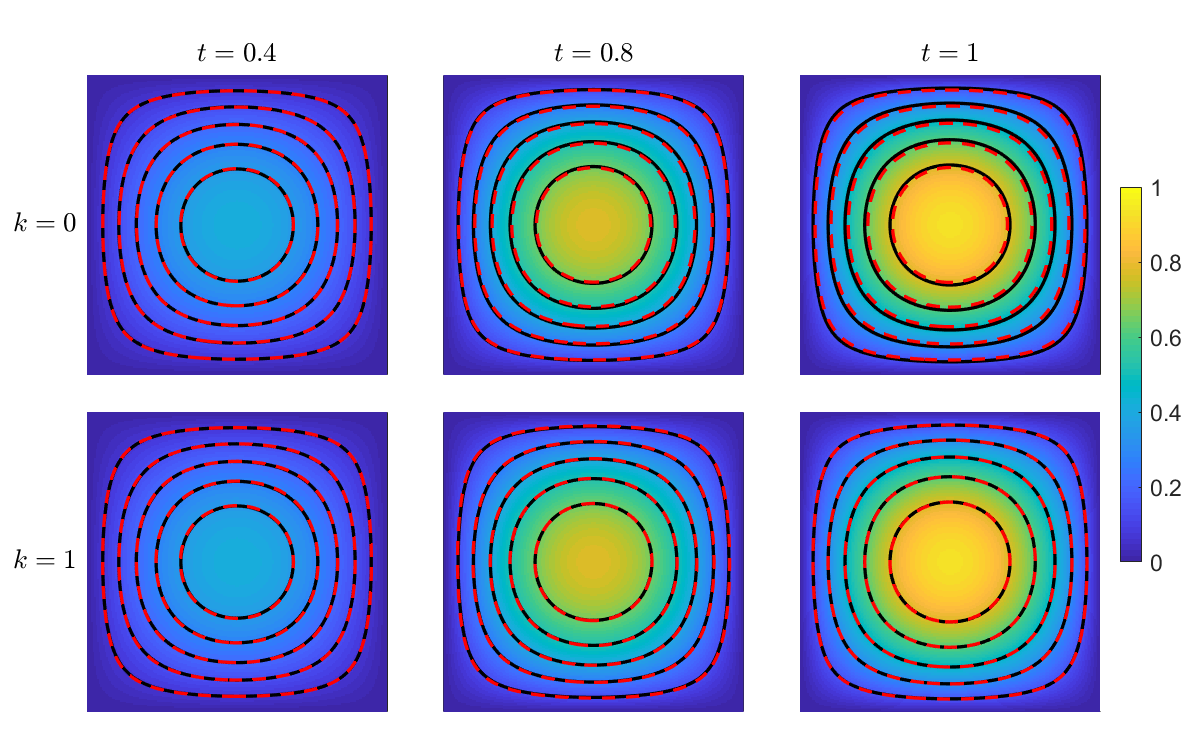}
    \caption{Example of using PPINN to solve the nonlinear diffusion-reaction equation.  First row ($k = 0$): First iteration, Second row ($k = 1$): Second iteration. Black solid line: Reference solution, which is obtained from the lattice Boltzmann method using a uniform grid of $x \times y \times t = 200 \times 200 \times 100$ \cite{meng2016localized}. Red dashed line: Solution from the PPINN.}
    \label{fig:reaction}
\end{figure}

\begin{figure}[h!]
    \centering
    \includegraphics[width=0.8\textwidth]{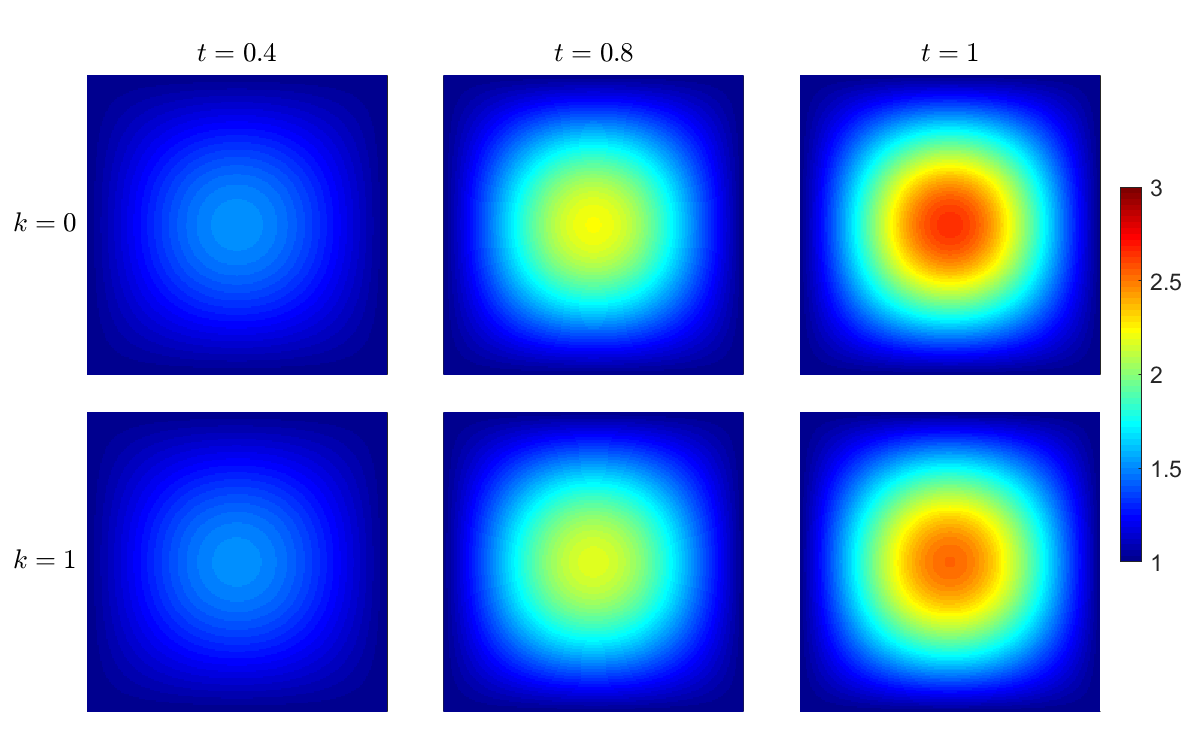}
    \caption{Distribution of the normalized diffusion coefficient ($D/D_0$).  First row ($k = 0$): First iteration, Second row ($k = 1$): Second iteration.}
    \label{fig:diffusion}
\end{figure}

We employ 2,000 random residual points to compute the $\mbox{MSE}_{R}$ in each subdomain for the coarse solver. We also employ 1,000 random points to compute the $\mbox{MSE}_{BC}$ for each boundary, and 2,000 random points to compute the $\mbox{MSE}_{IC}$. For each fine PINN, we use 5,000 random residual points for $\mbox{MSE}_{R}$, while the numbers of training points for the boundary and initial conditions are kept the same as those used in the coarse solver. It takes two iterations to meet the convergence criterion, i.e. $E_{tol} = 1\%$. The PPINN solutions as well as the distribution of the diffusion coefficient for each iteration at three representative times (i.e., $t = 0.4$, 0.8, and 1) are displayed in Figs. ~\ref{fig:reaction} and ~\ref{fig:diffusion}, respectively. We see that the solution at $t = 0.4$ agrees well with the reference solution after the first iteration, while the discrepancy between the PPINN and reference solutions increases with the time, which can be observed for $t = 0.8$ and 1 in the first row of Fig. \ref{fig:reaction}. This is reasonable because the error of the initial condition for each subdomain will increase with time. In addition, the solutions at the three representative times agree quite well with the reference solutions after the second iteration, as demonstrated in the second row in Fig.~\ref{fig:reaction}. All the above results again confirm the accuracy of the present approach. 

\begin{table}[htbp]
\centering
 \begin{tabular}{c|cccccc}
  \hline \hline
  Subdomains ($\#$) & NNs & $\mbox{MSE}_{R}$ & $\mbox{MSE}_{BC}$ & $\mbox{MSE}_{IC}$ & Optimizer & Learning rate \\ \hline
 1    &  $[20] \times 3$ & 200,000 & $10,000 \times 4$ & 2,000 & Adam, L-BFGS & $10^{-3}$   \\
 10    &  $[20] \times 2$ & 20,000 & $1,000 \times 4$ & 2,000 & Adam, L-BFGS & $10^{-3}$   \\
 20    &  $[16] \times 2$ & 10,000 & $500 \times 4$ & 2,000 & Adam, L-BFGS & $10^{-3}$   \\
 40    &  $[16] \times 2$ & 5,000 & $250 \times 4$ & 2,000 & Adam, L-BFGS & $10^{-3}$   \\
 50    &  $[16] \times 2$ & 400 & $200 \times 4$ & 2,000 & Adam, L-BFGS & $10^{-3}$   \\
  \hline \hline
 \end{tabular}
  \caption{\label{tab:reaction}Parameters used in the PPINN for modeling the two-dimensional diffusion-reaction system. The NNs are first trained using the Adam optimizer with a learning rate $10^{-3}$ until the loss is less than $10^{-3}$, then we utilize the L-BFGS-B to further train the NNs \cite{2019Raissi}. }
\end{table}

\begin{table}[htbp]
\centering
 \begin{tabular}{c|ccccc}
  \hline \hline
  &Subdomains ($\#$) & Iterations ($\#$) & $T_{total}$ (s) & $S$\\ \hline
 \multirow{5}{*}{PPINN (PINN)} 
 &1&  - & 27,627 & - \\
&10  &  2  & 6,230 & 4.3\\
&20  &  3  &  7,620 & 3.6 \\
&40  &  3 &  20,748 & 1.3 \\
&50 &  3 & $>$ 27,627 & - \\ \hline
  \multirow{5}{*}{PPINN (FDM)} 
  &1&  - & 27,627 & - \\
&10  &  2 & 5,421  & 5.1 \\
&20  &  2  & 2,472  & 11.2 \\
&40  &  3 & 1,535 & 18.0 \\
&50 &  3 & 1,620 & 17.1 \\                         
  \hline \hline
 \end{tabular}
  \caption{\label{tab:reaction_speedup} Speed-ups for using the PPINN with different coarse solvers to model the nonlinear diffusion-reaction system.   $T_{total}$ denotes the total computational time taken by the PPINN, and $S$ is the speed-up ratio.}
\end{table}

We further investigate the speed-up of the PPINN using the PINN as coarse solver. Here $T = 10$, $A = 0.1146$, and $R = 0.5$,  and we use five different subdomains, i.e., 1, 10, 20, 40, and 50. The parallel code is run on multiple CPUs (Intel Xeon E5-2670). The total number of the residual points in the coarse solver is 20,000, which is uniformly divided into $N$ subdomains. The number of training points for the  boundary condition on each boundary is 10,000, which is also uniformly assigned to each subdomain. In addition,  2,000 randomly sampled points are employed for the initial condition. The architecture and other parameters (e.g., learning rate, optimizer, etc) for the CG PINN in each subdomain are the same as the fine PINN, which are displayed in Table \ref{tab:reaction}. The speed-up ratios for the representative cases are shown in Table \ref{tab:reaction_speedup}. We notice that the speed-up does not increase monotonically with the number of subdomains. On the contrary, the speed-up decreases with the number of subdomains. This result suggests that the cost for the CG PINN is not only related to the number of training data but is also affected by the number of hyperparameters. The total hyperparameters in the CG PINNs increases with the number of subdomains, leading to the walltime for the PPINN to increase with the number of subdomains. 

To validate the above hypothesis, we replace the PINN with the finite difference method~\cite{meerschaert2006finite} (FDM, grid size: $x \times y = 20 \times 20$, time step $\delta_t = 0.05$) for the coarse solver, which is much more efficient than the PINN. The walltime for the FDM in each subdomain is negligible compared to the fine PINN. As shown in Table \ref{tab:reaction_speedup}, the speed-ups are now proportional to the number of subdomains as expected.

\section{Summary and Discussion}\label{sec:summary}

Our overarching goal is to develop {\em data-driven} simulation capabilities that go beyond the traditional data assimilation methods by taking advantage of the recent advances in deep learning methods. Physics-informed neural networks~(PINNs)~\cite{2019Raissi} play exactly that role but they become computationally intractable for very long-time integration of time-dependent PDEs. Here, we address this issue for first time by proposing 
a parallel-in-time PINN (PPINN). In this approach, two different PDEs are solved by the coarse-grained (CG) solver and the fine PINN, respectively. In particular, the CG solver can be any efficient numerical method since the CG solver solves a surrogate PDE, which can be viewed as either a simplified form of the exact PDE or it could be a reduced-order model or any other surrogate model, including an offline pre-trained PINN although this was not pursued here. The solutions of the CG solver are then serving as initial conditions for the fine PINN in each subdomain. The fine PINNs can then run in parallel independently based on the provided initial conditions, hence significantly reduce their training cost. In addition, it is worth mentioning that the walltime for the PINN in each subdomain at $k$-{th} ($k > 1$) iteration is negligible compared to the first iteration for the fine PINN, which leads to a  superlinear speed-up of PPINNs assuming an efficient CG solver is employed.

The convergence of the PPINN algorithm is first validated for deterministic and stochastic ODE problems. In both cases, we employ a simplified ODE in the coarse solver to supervise the fine PINNs. The results demonstrate that the PPINN can converge in a few iterations as we decompose the 
time-domain into 10 uniform subdomains. Furthermore, superlinear speed-ups are achieved for both cases. Two PDE problems, i.e., the one-dimensional Burgers equation and the two-dimensional diffusion-reaction system with nonlinear diffusion coefficient, are further tested to validate the present algorithm. For the 1D case, the simplified PDE is also a Burgers equation but with an increased viscosity, which yields much smoother solution and thus can be solved more efficiently. In the diffusion-reaction system, we use a constant diffusion coefficient instead of the nonlinear one in the CG solver, which is also much easier to solve compared to the exact PDE. The results showed that the fine PINNs can converge in only a few iterations based on the initial conditions provided by the CG solver in both cases. Finally, we also demonstrate that  a superlinear speed-up can be obtained for the two-dimensional case if the CG solver is efficient.

We would like to point out another advantage of the present PPINN. In general, GPUs have very good computational efficiency in training DNNs with big-data sets. However, when one works on training a DNN with small-data sets, CPUs may have comparable performance to GPUs. The reason is that we need to transfer the dataset from the CPU memory to the GPU, and then transfer it back to the CPU memory, which can be even more time consuming than the computational process for small dataset \cite{sanders2010cuda}. For example, given a DNN with architecture of $[2]+[20]*3+[1]$, the wall time of training the PINN (1D Burgers equation) for 10,000 steps is about 100 seconds on an Intel Xeon CPU (E5-2643) and 172 seconds on a NVIDIA GPU (GTX Titan XP) for a training data set consisting of 5,000 points. 
Because the proposed PPINN approach is able to break a big-data set into multiple small-data sets for training, it enables us to train a PINN with big-data sets on CPU resources if one does not have access to sufficient GPU resources.   

The concept of PPINN could be readily extended to domain decomposition of physical problems with large spatial databases by drawing an analog of multigrid/multiresolution methods~\cite{1998Beylkin}, where a cheap coarse-grained (CG) PINN should be constructed and be used to supervise and connect the PINN solutions of sub-domains iteratively. Moreover, in this prediction-correction framework, the CG PINN only provides a rough prediction of the solution, and the equations encoded into the CG PINN can be different from the equations encoded in the fine PINNs. Therefore, PPINN is able to tackle multi-fidelity modeling of inverse physical problems~\cite{2019Meng}. 
Both topics are interesting and would be investigated in future work.

\section*{Acknowledgements}
This work was supported by the DOE PhILMs project~DE-SC0019453 and the DOE-BER
grant~DE-SC0019434. This research was conducted using computational resources and services at the Center for Computation and Visualization, Brown University. Z.\ Li would like to thank Dr.\ Zhiping Mao, and Dr.\ Ansel L Blumers for helpful discussions. 

\bibliographystyle{unsrt}
\bibliography{references}

\begin{thebibliography}{10}

\bibitem{2019Raissi}
M.~Raissi, P.~Perdikaris, and G.~E. Karniadakis.
\newblock Physics-informed neural networks: A deep learning framework for
  solving forward and inverse problems involving nonlinear partial differential
  equations.
\newblock {\em J. Comput. Phys.}, 378:686--707, 2019.

\bibitem{2018Hodas}
N.~O. Hodas and P.~Stinis.
\newblock Doing the impossible: Why neural networks can be trained at all.
\newblock {\em Front. Psychol.}, 9(1185), 2018.

\bibitem{2018Kurth}
T.~Kurth, S.~Treichler, J.~Romero, M.~Mudigonda, N.~Luehr, E.~Phillips,
  A.~Mahesh, M.~Matheson, J.~Deslippe, M.~Fatica, Prabhat, and M.~Houston.
\newblock Exascale deep learning for climate analytics.
\newblock {\em arXiv preprint arXiv:1810.01993}, 2018.

\bibitem{2019Lew}
J.~{Lew}, D.~A. {Shah}, S.~{Pati}, S.~{Cattell}, M.~{Zhang}, A.~{Sandhupatla},
  C.~{Ng}, N.~{Goli}, M.~D. {Sinclair}, T.~G. {Rogers}, and T.~M. {Aamodt}.
\newblock Analyzing machine learning workloads using a detailed {GPU}
  simulator.
\newblock In {\em 2019 IEEE International Symposium on Performance Analysis of
  Systems and Software (ISPASS)}, pages 151--152, 2019.

\bibitem{2019You}
Y.~You, Z.~Zhang, C.~Hsieh, J.~Demmel, and K.~Keutzer.
\newblock Fast deep neural network training on distributed systems and cloud
  {TPUs}.
\newblock {\em {IEEE} Transactions on Parallel and Distributed Systems}, pages
  1--14, 2019.

\bibitem{2018Bottou}
L.~Bottou, F.~Curtis, and J.~Nocedal.
\newblock Optimization methods for large-scale machine learning.
\newblock {\em SIAM Review}, 60(2):223--311, 2018.

\bibitem{2014Kingma}
D.~P. Kingma and J.~Ba.
\newblock Adam: A method for stochastic optimization.
\newblock {\em arXiv preprint arXiv:1412.6980}, 2014.

\bibitem{2011Duchi}
J.~Duchi, E.~Hazan, and Y.~Singer.
\newblock Adaptive subgradient methods for online learning and stochastic
  optimization.
\newblock {\em J. Mach. Learn. Res.}, 12(Jul):2121--2159, 2011.

\bibitem{2019Michoski}
C.~Michoski, M.~Milosavljevic, T.~Oliver, and D.~Hatch.
\newblock Solving irregular and data-enriched differential equations using deep
  neural networks.
\newblock {\em arXiv preprint arXiv:1905.04351}, 2019.

\bibitem{2019Doan}
N.~A.~K. Doan, W.~Polifke, and L.~Magri.
\newblock Physics-informed echo state networks for chaotic systems forecasting.
\newblock In {\em {ICCS 2019 - International Conference on Computational
  Science}}, Faro, Portugal, 2019.

\bibitem{2019Mattheakis}
M.~Mattheakis, P.~Protopapas, D~Sondak, M.~Di~Giovanni, and E.~Kaxiras.
\newblock Physical symmetries embedded in neural networks.
\newblock {\em arXiv preprint arXiv:1904.08991}, 2019.

\bibitem{1994Dissanayake}
M.~W. M.~G. Dissanayake and N.~Phan-Thien.
\newblock Neural-network-based approximations for solving partial-differential
  equations.
\newblock {\em Comm. Numer. Meth. Eng.}, 10(3):195--201, 1994.

\bibitem{yang2018physics}
L.~Yang, D.~Zhang, and G.~E. Karniadakis.
\newblock Physics-informed generative adversarial networks for stochastic
  differential equations.
\newblock {\em arXiv preprint arXiv:1811.02033}, 2018.

\bibitem{2018Zhang}
D.~Zhang, L.~Lu, L.~Guo, and G.~E. Karniadakis.
\newblock Quantifying total uncertainty in physics-informed neural networks for
  solving forward and inverse stochastic problems.
\newblock {\em arXiv preprint arXiv:1809.08327}, 2018.

\bibitem{zhang2019learning}
D.~Zhang, L.~Guo, and G.~E. Karniadakis.
\newblock Learning in modal space: Solving time-dependent stochastic {PDE}s
  using physics-informed neural networks.
\newblock {\em arXiv preprint arXiv:1905.01205}, 2019.

\bibitem{yang2019adversarial}
Y.~Yang and P.~Perdikaris.
\newblock Adversarial uncertainty quantification in physics-informed neural
  networks.
\newblock {\em J. Comput. Phys.}, 394:136--152, 2019.

\bibitem{pang2018fpinns}
G.~Pang, L.~Lu, and G.~E. Karniadakis.
\newblock f{PINN}s: Fractional physics-informed neural networks.
\newblock {\em arXiv preprint arXiv:1811.08967}, 2018.

\bibitem{2002Maday}
Y.~Maday and G.~Turinici.
\newblock A parareal in time procedure for the control of partial differential
  equations.
\newblock {\em C. R. Math.}, 335(4):387--392, 2002.

\bibitem{2019Blumers}
A.~Blumers, Z.~Li, and G.~E. Karniadakis.
\newblock Supervised parallel-in-time algorithm for long-time {L}agrangian
  simulations of stochastic dynamics: Application to hydrodynamics.
\newblock {\em J. Comput. Phys.}, 393:214--228, 2019.

\bibitem{2014Livni}
R.~Livni, S.~Shalev-Shwartz, and O.~Shamir.
\newblock On the computational efficiency of training neural networks.
\newblock In {\em Advances in neural information processing systems}, pages
  855--863, 2014.

\bibitem{meng2016localized}
X.~Meng and Z.~Guo.
\newblock Localized lattice {B}oltzmann equation model for simulating miscible
  viscous displacement in porous media.
\newblock {\em Int. J. Heat Mass Tran.}, 100:767--778, 2016.

\bibitem{meerschaert2006finite}
M.~M. Meerschaert, H.-P. Scheffler, and C.~Tadjeran.
\newblock Finite difference methods for two-dimensional fractional dispersion
  equation.
\newblock {\em J. Comp. Phys.}, 211(1):249--261, 2006.

\bibitem{sanders2010cuda}
J.~Sanders and E.~Kandrot.
\newblock {\em CUDA by example: an introduction to general-purpose GPU
  programming}.
\newblock Addison-Wesley Professional, 2010.

\bibitem{1998Beylkin}
G.~Beylkin and N.~Coult.
\newblock A multiresolution strategy for reduction of elliptic {PDEs} and
  eigenvalue problems.
\newblock {\em Appl. Comput. Harmon. Anal.}, 5(2):129--155, 1998.

\bibitem{2019Meng}
X.~Meng and G.~E. Karniadakis.
\newblock A composite neural network that learns from multi-fidelity data:
  Application to function approximation and inverse {PDE} problems.
\newblock {\em arXiv preprint arXiv:1903.00104}, 2019.

\end{thebibliography}

\end{document}